
\hsize 5.50in
\vsize 8.5in
\parskip 0pt
\def\({[}
\def\){]}

\def\rjustline#1{\line{\hss#1}}

\def\rarrow{\rightarrow}
\parindent 0.5in

\overfullrule 0pt
\rjustline{WIS-92/105/Dec-PH}
\baselineskip 12pt
\bigskip
PARADOXES IN STRANGE QUARK CONTRIBUTIONS TO

HYPERON SPINS
\vskip0.4in

Harry.~J.~LIPKIN

Department of Nuclear Physics, Weizmann Institute of Science

Rehovot,  76100

\centerline {and}

School of Physics and Astronomy

Raymond and Beverly Sackler Faculty of Exact Sciences

Tel Aviv University, Tel Aviv,

ISRAEL
\vskip0.5in

{\noindent \us{ABSTRACT}
\vskip0.1in

Data for magnetic moments and semileptonic decays show disagreements
between experimental values and theoretical predictions not easily
explained by simple models for $\Lambda$ and $\Sigma$ hyperons.

\vskip0.1in

{\noindent 1. \us{ Introduction - A Strong Disagreement with SU(6)}}
\vskip0.1in

Baryon magnetic moment data
agree with the naive SU(6) quark model at the 15\% level
but are now good to a few per cent\REF{\Heller}{K.
Heller, these proceedings} $[{\Heller}]$ and may provide clues to new
physics. Detailed models have too many free parameters for significant
tests against 8 or 9 experimental numbers. Instead we suggest
investigating: 1) General features like
the roles of the strange quarks in the $\Lambda$,
$\Sigma$ and proton;
2) New kinds of data like
hyperon beam investigations of Primakoff excitation of $\Sigma^*$ and
$\Xi^*$ resonances\REF{\LIPHYP}{Harry J. Lipkin, 1973, \us {Phys. Rev.}
D7, 846.}
\REF{\MURLIP}{Harry J. Lipkin and Murray A. Moinester, 1992,
\us {Phys. Lett.} B287, 179.}$[{\LIPHYP,\MURLIP}]$;
3) Lattice QCD results for spin structure of baryons
\REF{\LATTICE}{D.B. Leinweber, R.M. Woloshyn, and T. Draper, 1991
\us {Phys. Rev.} 43D, 1659.}$[{\LATTICE}]$
or exotic multiquark states like H dibaryon and
Pentaquark\REF{\PHPROD}{Murray A. Moinester, Carl B. Dover and Harry J. Lipkin,
1992, \us {Phys. Rev.} C46 1082.}$[{\PHPROD}]$.
This talk focuses on the contributions
$ \Delta s (\Lambda)$ and $ \Delta s (\Sigma)$
of the strange quark in the $\Lambda$ and $\Sigma$ to the
hyperon spin.

The SU(6) values $ \Delta s (\Lambda)_{SU(6)} = 1$ and
$ \Delta s (\Sigma)_{SU(6)} = -1/3$
lead to predictions in disagreement with experiment\REF{\PDG}{Particle
Data Group, 1990, \us{Phys. Lett.} B239, 1.}$[{\PDG}]$
in opposite directions for weak decays and magnetic moments.
Semileptonic decays give too large a value for the ratio
$ \Delta s (\Sigma)/ \Delta s (\Lambda)$; magnetic moments give too low.
 $$
 -1/3 = {{\Delta s (\Sigma)_{SU(6)}}\over{\Delta s (\Lambda)_{SU(6)}}} =
 {{(G_A/G_V)_{\Sigma^-\rightarrow n}}\over
{(G_A/G_V)_{\Lambda \rarrow p}}} =   -0.473 \pm 0.026
\eqno(YY1a)   $$
 $$
 -1/3 = {{\Delta s (\Sigma)_{SU(6)}}\over{\Delta s (\Lambda)_{SU(6)}}} =
{{\mu_{\Sigma^{+}} + 2\mu_{\Sigma^{-}}}\over
{3\mu_{\Lambda}}}= - 0.06 \pm 0.02 \eqno(YY1b)   $$
$$ 1 = {{(G_A/G_V)_{\Lambda \rarrow p}}\over
{(G_A/G_V)_{\Sigma^-\rightarrow n}}}\cdot
{{\mu_{\Sigma^{+}} + 2\mu_{\Sigma^{-}}}\over{\mu_{\Lambda}}}=
0.12  \pm 0.04  \eqno(QQ2)  $$
This enormous discrepancy by a factor of $8 \pm 2$ is not easily overcome.
The excellent agreement obtained\REF{\DGG}{
A. De Rujula, H. Georgi and S.L. Glashow, 1975 \us {Phys. Rev.} D12 147.}
$[{\DGG}]$ for $\mu_{\Lambda}$ by assuming
$ \Delta s (\Lambda) = 1$
supports naive SU(6) for the $\Lambda$; $\mu_{\Sigma^\pm}$
disagree with SU(6). The excellent
agreement with SU(6) for
$(G_A/G_V)_{\Sigma^-\rightarrow n}$ supports naive SU(6)
for the $\Sigma$; the $\Lambda$ decay disagrees with SU(6).
These $\Lambda$ - $\Sigma$ disagreements are very general;
they do not assume flavor SU(3) symmetry and
only consider states having nearly the same masses.
We now examine the underlying physics.

\vskip0.1in
{\noindent 2. \us{ A General Treatment of $\Sigma$ Magnetic Moments}
\vskip0.1in

We define $\mu_f$ as the most general one-body
operator acting only on quarks and antiquarks of flavor $f$,
including orbital and sea contributions. We write the $\Sigma$ moments
as the expectation value of the sum of three terms, one for each flavor,
each proportional to the quark charge, in the most general wave functions
satisfying isospin invariance.
$$ \mu_{\Sigma^\pm} =
\bra {\Sigma^{\pm}}
{2\over 3}\cdot\mu_u -{1\over 3}\cdot\mu_d -{1\over 3}\cdot\mu_s
\ket{\Sigma^{\pm}} =
\bra {\Sigma^{\mp}}
{2\over 3}\cdot\mu_d -{1\over 3}\cdot\mu_u -{1\over 3}\cdot\mu_s
\ket{\Sigma^{\mp}}
 \eqno(YY3)   $$
 $$
{{\mu_{\Sigma^{+}} + 2\mu_{\Sigma^{-}}}\over
{3\mu_{\Sigma^{+}}}}=
{{\bra {\Sigma^{+}} \mu_s^V + \mu_s^S - \mu_d^S
\ket{\Sigma^{+}}}\over
{3\mu_{\Sigma^{+}}}}= 0.015 \pm 0.005
\eqno(QQ4)   $$
where we have separated valence and sea contributions
denoted by $\mu_f^V$ and $\mu_f^S$. Thus
the contribution to $\mu_\Sigma$ of the valence strange quark is either
anomalously low, only 1.5\% of $\mu_{\Sigma^+}$ as previously
noted\REF{\LIPMAG}{Harry J. Lipkin, 1984, \us{Nucl. Phys.} B241, 477.}
$[{\LIPMAG}]$, or mysteriously canceled by
an appreciable contribution from SU(3) breaking in the sea.
This result is more general than
previous similar treatments\REF{\KARL}{G. Karl, 1992,
\us{Phys. Rev.} D45
247.} $[{\KARL}]$.

\vskip0.1in
{\noindent 3. \us{ A General Treatment of Hyperon Semileptonic Decays}
\vskip0.1in

A previous analysis of semileptonic
n, $\Lambda$, $\Sigma$ and $\Xi$ decay data\REF{\SPINPF}{Harry J.
Lipkin, 1992, in \us{The Vancouver Meeting Particles and Fields '91}
Proceedings of PF91, Vancouver, Canada August 18-22 (1991)
Edited by David Axen, Douglas Bryman and Martin Comyn,
World Scientific,  Singapore (1992) p. 603.}
$[{\SPINPF}]$ showed the contrast between neutron and $\Lambda$ decays
in strong disagreement with simple SU(6) predictions but both smaller by
the same factor, and $\Sigma$ decays in striking agreement with SU(6),
while large errors left $\Xi$ data within two standard deviations of
both SU(6) predictions and consistency with $n$ and $\Lambda$

The $\Lambda$ and $\Sigma$ decays are very simply described as
a valence $s \rightarrow u$ quark transition with all remaining degrees
of freedom including any combination of valence $d$ quarks,
$q \bar q$ pairs, gluons and orbital angular momenta remaining inert
spectators. Any $s \bar s$ pairs in the nucleon have a
completely different momentum spectrum from the valence quarks and can
play no active role in weak decays at zero momentum transfer.
In the baryon rest frame the spectator total angular momentum has
only two allowed values, zero and one and
the $\Sigma$, $\Lambda$ and neutron wave functions can be written
$$ \ket{B} = {\rm cos} \theta^B
\ket{S_1^B; \,f} + {\rm sin} \theta^B
\ket{S_o^B; \,f}
\eqno (QQ5)$$
where $S_o^B$ and $S_1^B$,
denote the spectator states in the baryon $B$ wave function with
angular momentum zero and one respectively
and $\ket{S_k^B; \,f}$ denotes a state in baryon $B$ in which the angular
momenta of the spin-k spectator and an active quark of flavor $f$ are
coupled to total angular momentum 1/2 and
$ \theta^B $
denotes a ``mixing
angle" parameter defining the relative contribution of the two terms.

In both current and constituent quark models
the decay is an $s \rightarrow u$ transition
with or without spin flip while leaving the spectators unchanged.
The transition matrix element
factorizes into the product of a weak $s \rightarrow u$ matrix element
and an overlap factor between the final baryon state and the
state produced by the flavor and spin change of the active quark.
The value of $G_A/G_V$ for
the transition is measured experimentally by the ratio of the flip to
nonflip amplitudes.
$$ (G_A/G_V)_{B_i \rarrow B_f} =
{{\bra{u\downarrow}H_{weak}\ket{s\uparrow}}\over
{\bra{u\uparrow}H_{weak}\ket{s\uparrow}}}\cdot
$$
$$ \cdot  {{\sin \theta^i \bra{B_f\downarrow}{u\downarrow S_0^i \rangle} -(1/3)
\cos \theta^i \bra{B_f\downarrow}{(S_1^i;u)_{\downarrow} \rangle}}\over
{\sin \theta^i \bra{B_f\uparrow}{u\uparrow S_0^i \rangle} +
\cos \theta^i \bra{B_f\uparrow}{(S_1^i;u)_{\uparrow} \rangle}}}   \eqno(QQ6)
 $$
The two overlap factors $\bra{B_f\downarrow}
{u\downarrow S_0^i \rangle}$ and $\bra{B_f\uparrow}{u\uparrow  S_0^i \rangle}$
are equal by rotational invariance, completely independent of the structure of
the final state baryon $B_f$, and similarly for
$\bra{B_f\downarrow}{(S_1^i;u)_{\downarrow} \rangle}$ and
$ \bra{B_f\uparrow}{(S_1^i;u)_{\uparrow} \rangle}$.
The ratio of the matrix elements of $H_{weak}$ can be
interpreted as the value of $ (G_A/G_V) $ at the quark level. Thus
$$  (G_A/G_V)_{B_i \rarrow B_f} =
-{1\over 3} \cdot (g_A/g_V)_{s \rarrow u} \cdot {{ 1 - 3 \xi}\over
{ 1 + \xi}} \eqno (QQ7a) $$
where
$$ \xi = {\rm tan} \theta^i\cdot
 {{\bra{B_f\downarrow}{u\downarrow S_0^i \rangle}}\over
{\bra{B_f\downarrow}{(S_1^i;u)_{\downarrow} \rangle}}}
\eqno(QQ7b)  $$

The SU(6) results for
$ (G_A/G_V)_{(\Sigma^-\rightarrow n)}$ and $ (G_A/G_V)_{\Lambda \rarrow p}$
are seen to follow only from assuming that the spectator degrees of freedom
in the hyperons are coupled to the SU(6) values $1$ and $0$ respectively,
to give ($ \theta^\Sigma = 0$;
$ \theta^\Lambda = 90^o$)
with no assumptions about the structures of the spectators nor
the final state nucleon.
$$  (G_A/G_V)_{B_i \rarrow B_f} = -(1/3)
(g_A/g_V)_{s \rarrow u}
 {\rm \ \  \ \ if \sin \theta^i =0}
\eqno(QQ8a)  $$
$$  (G_A/G_V)_{B_i \rarrow B_f} = (g_A/g_V)_{s \rarrow u}
 {\rm \ \  \ \ if \cos \theta^i =0}
\eqno(QQ8b)  $$
This highlights the difference between the agreement
of $ (G_A/G_V)_{(\Sigma^-\rightarrow n)}$ with SU(6) and
the disagreement of $ (G_A/G_V)_{\Lambda \rarrow p}$.

The $\Lambda \rarrow p$ decay is particularly simple in any model where the
strange quark carries the spin of the baryon and the
other degrees of freedom including the nonstrange quarks are
coupled to angular momentum zero.
Then ${\rm cos} \theta^\Lambda= 0$, eq. (10b) is valid and comparison
with experiment
$[{\PDG}]$
gives
$$ (G_A/G_V)_{\Lambda \rarrow p} =
(g_A/g_V)_{s \rarrow u}
=  0.718 \pm 0.015
\eqno(QQ9)  $$
Thus either the the value of
$ (G_A/G_V) $ at the quark level is reduced to the experimental
value 0.72 or the $\Lambda$ wave function with all the spin carried by
the strange quark is not valid.

The $\Sigma^-\rightarrow n$ decay is uniquely simple in all models
where a single ``active" $s$ quark in the $\Sigma^-$ turns
into a $u$ quark in the neutron, with all the remaining degrees of
freedom including all the $d$ quarks remaining inert spectators.
The value of $ (G_A/G_V)_{(\Sigma^-\rightarrow n)}$ is
given by eq. (QQ7a) for the most general wave functions (QQ5)
with $\xi$ given by
substituting (QQ5) into (QQ7b). Comparing this value with experiment
$[{\PDG}]$ gives
$$ (G_A/G_V)_{(\Sigma^-\rightarrow n)} =
-{1\over 3} \cdot (g_A/g_V)_{s \rarrow u} \cdot {{ 1 - 3 \xi}\over
{ 1 + \xi}} =  -0.340 \pm 0.017 \eqno (QQ10a) $$
$$ \xi = {\rm tan} \theta^\Sigma  {\rm tan} \theta^n
\cdot {{\langle S_o^n \ket{S_o^\Sigma}  }\over
{\langle S_1^n\ket{S_1^\Sigma}}}
\eqno (QQ10b) $$

In the SU(3) symmetry limit $\xi > 0$ and is expected
to remain positive even when SU(3) is broken.
Thus the
result (QQ10a) requires $(g_A/g_V)_{s \rarrow u} \geq 1$, while the
experimental result$[{\PDG}]$ $G_A/G_V  = 1.259 \pm 0.004$ for the
neutron beta decay is in strong disagreement with the SU(6) prediction
$G_A/G_V  =  5/3$ implies $(g_A/g_V)_{d \rarrow u} \approx 3/4$.

An analogous treatment is not possible for $ (G_A/G_V)_{\Lambda \rarrow p}$,
where there are two active $u$ quarks in the proton, and the wave function
$B_f$ cannot be written in the form (QQ5) which has only one active quark.
The neutron decay is even more complicated, since there are two active $d$
quarks in the neutron as well as two active $u$ quarks in the proton,
either pair of active quark spins can be coupled to spin zero or spin one,
and there are many more free parameters depending upon the wave functions in
the expression for $  (G_A/G_V)_{(n\rightarrow p)}$.

Why is the $\Sigma$ different from all other baryons?
The large value in agreement of experiment with SU(6)
$ (G_A/G_V)_{(\Sigma^-\rightarrow n)} =-(1/3)$
for the simplest weak decay where the prediction is least dependent upon
wave function structure implies that the spin projection of the strange
valence quark in the $\Sigma$ is antiparallel to the hyperon spin and has
the largest possible value. Yet the contribution of this strange quark
spin to the $\Sigma$ magnetic moment seems to be mysteriously suppressed
by a large factor or cancelled by some other unknown contribution.
There is no such suppression observed in the contributions to nucleon
magnetic moments of the $d$ quark in the proton and the $u$ quark in
the neutron, which are directly related by $SU(3)$ to the strange quark
contribution in the $\Sigma$, and all other weak decays seem to have
$ (G_A/G_V) $ suppressed by a factor of the order of $3/4$.

\vskip0.1in
{\noindent 4. \us{ Additional Input Obtainable from Other Electromagnetic
Properties}
\vskip0.1in

Additional insight may be obtained from
experimental data on other electromagnetic properties; e.g. charge radii,
polarizabilities and $B \rightarrow B^*$ transitions$[{\MURLIP}]$.
Any difference
between the electromagnetic properties of the $\Sigma^+$ and proton can only
arise from flavor SU(3) symmetry breaking. The two $\Sigma$ states are isospin
mirrors but have very different electrical quark
structures. The $\Sigma^+$, like the nucleon
and $\Xi^o$, has valence quarks of two flavors having + and -
electric charge. External fields act in
opposite directions on the two flavors and rotate spins in opposite directions,
thereby producing internal excitation. The $\Sigma^-$ and
$\Xi^-$, have three valence quarks all with charge -1/3.
External fields act in the
same direction on all three and rotate spins in the same
direction, producing no internal excitation in the $SU(3)$ symmetry
limit, as in the well-known $SU(3)$ U-spin selection rule$[{\LIPHYP}]$
${\Gamma({\Sigma^{-}}  \rightarrow {\Sigma^{*-}})}=
{\Gamma({\Xi^{-}}  \rightarrow {\Xi^{*-}})}= 0$.
Broken-SU(3) sum rules $[{\MURLIP}]$
have been derived
under the assumption that the contributions of the two
quarks of the same flavor in nucleons and $\Sigma$'s
are not changed by SU(3) symmetry breaking; e.g.
  $$  \sqrt{\Gamma({\Sigma^{-}}  \rightarrow {\Sigma^{*-}})}=
  \sqrt{\Gamma({\Sigma^{+}}  \rightarrow {\Sigma^{*+}})}
-  \sqrt{\Gamma(N \rightarrow \Delta)}
\eqno(QQ11a)   $$
$$
\langle r_c^2 \rangle_{\Sigma^{+}} + \langle r_c^2 \rangle_{\Sigma^{-}}
= 2 (\langle r_c^2 \rangle_{\Sigma^{+}} - \langle r_c^2 \rangle_p)
- \langle r_c^2 \rangle_n
\eqno(QQ11b)  $$
where $\langle r_c^2 \rangle_B$ denotes the mean square charge radius of baryon
B.
The sum rule (QQ11a) should hold separately for the $M1$ and $E2$
octet-decuplet transitions.

\refout
\end